\documentclass[article,pre,10pt,twocolumn,longbibliography]{revtex4-1}
\usepackage{microtype}
\usepackage{graphicx, svg}
\usepackage{amsmath}
\usepackage{tcolorbox}
\graphicspath{{./images/}}
\usepackage[colorlinks,linkcolor=blue,urlcolor=blue,citecolor=blue, pdfencoding=auto, psdextra]{hyperref}
\usepackage[english]{babel}
\usepackage{ragged2e}
\usepackage{caption}
\usepackage[labelfont=bf, textfont=small]{caption}
\usepackage{subcaption}
\usepackage{epsfig,amsopn}
\usepackage{graphicx}
\usepackage{bm,bbm,amsmath,amssymb,empheq,physics,cancel}
\usepackage{natbib}
\usepackage{braket}
\usepackage{float}
\usepackage{wasysym}
\usepackage{enumerate}
\usepackage{comment}
\allowdisplaybreaks[1]
\usepackage[normalem]{ulem} 
\def\be{\begin{equation}}
\def\ee{\end{equation}}
\def\bse{\begin{subequations}}
\def\ese{\end{subequations}}
\def\bcs{\begin{cases}}
\def\ecs{\end{cases}}
\def\bea{\begin{eqnarray}}
\def\eea{\end{eqnarray}}

\newcommand{\si}{s_i}

\newcommand{\Av}[1]{\mathbb{E}_{\vv{\xi}}\!\left[#1\right]}
\newcommand{\Avm}[1]{\mathbb{E}_{\xi,z}\!\left[#1\right]}

\newcommand{\Aref}[1]{Appendix}%

\newcommand{\opunit}{\textrm{1}\kern-0.22em\textrm{l}}

\newcommand{\xiu}{\xi_i^\mu}

\usepackage{makeidx}
\newcommand{\vv}[1]{\bm{#1}}
\begin{document}
\title{Free energy landscape of Dense Associative Memory}

\author{{\normalsize{}Sumedha,$^{1, 2}$}
{\normalsize{}}}
\email{sumedha@niser.ac.in}

\author{{\normalsize{}Abhishek Singh$^{1, 2}$}
{\normalsize{}}}

\affiliation{\noindent $^{1}$School of Physical Sciences, National Institute of Science Education and Research, Bhubaneswar, P.O. Jatni, 752050, India}
\affiliation{\noindent $^{2}$Homi Bhabha National Institute, Training School Complex, Anushakti Nagar 400094, India}

\begin{abstract}
Using large deviations theory, we solve and obtain a general expression for the free energy functional for a broad class of associative memories, including dense associative memories. We illustrate the method by reproducing classical results for the Hopfield model. For a finite number of patterns, we derive the temperature-dependent free energy functional for dense associative memories featuring polynomial interactions and Log-Sum-Exponential (LSE) activation. We also obtain the disorder-averaged ground-state energy landscape of these systems in the extensive limit. Our analytical framework reveals how memory retrieval depends on the initial state in higher-order dense networks, and gives the exact error-free threshold for the LSE model. This method provides a systematic procedure for analyzing diverse, complex architectures in associative memory.
\end{abstract}

\maketitle
Dense Associative Memories (DenseAMs) are energy-based neural architectures that vastly surpass traditional Hopfield networks by using higher-than-quadratic neuron interactions to expand memory storage capacity  \cite{krotov1,gardner,hopfield}. In these networks, stored patterns represent low-energy configurations, allowing the system to perform powerful error correction by mapping partial or corrupted inputs back to the correct pattern. This process closely mirrors the biological error-correction mechanisms of the human brain, which makes it possible for it to retrieve memories from incomplete information \cite{anderson}. Because of this high-capacity retrieval and error-correction,  DenseAMs have become useful tools in areas like modern deep learning, generative AI, pattern recognition, and transformer architectures  \cite{krotov2,simon,yampolskaya,transformer}.

In his seminal work, Hopfield mapped neuronal states to Ising spins and synaptic weights to the Hebb rule \cite{hebb}. This established a spin-glass-like model where specific configurations serve as stored memories. As a result, the  model was successfully solved using the replica method \cite{amit1,amit2}.

In associative memory (AM) models, gradient descent dynamics drives the retrieval of stored memories by guiding the system towards the nearest local minimum on a free-energy landscape. The recent work in DenseAMs though have primarily concentrated on energy dynamics rather than the full free-energy framework. Unlike the Hopfield model, the free energy  functional for these models is not known in closed form. In this paper, we derive the free energy landscape for a general class of associative memories and use it to study higher  DenseAMs \cite{krotov1,ramsauer,demi,lucibello}. We develop a method of quenched average that makes use of the tilted property  of the  measure of exponential functions to obtain the rate function. We had  earlier used similar methods to study random field quenched disorder for ferro-magnets \cite{sumedha-sushant,soheli-sumedha,sumedha-barma}. 

The main contributions of our work are as follows: We find a unifying framework to study different AM models like polynomial DenseAM and LSE model of associative memory. It  solves a broad class of AM models defined via Eq. \ref{eq2} for finite memories in the thermodynamic limit  straightforwardly. For extensive pattern limit we obtain the disorder averaged ground state energy landscape at zero temperature for retrieval of a memory by taking unaligned memories overlap to be of order $O(1/\sqrt{N})$. This gives the landscape of the gradient descent dynamics which connects the dynamical accessibility to the metastability of the underlying landscape. We show that the LSE cost function essentially changes the nature of the landscape compared to polynomial DenseAMs by tilting the landscape. As a result, LSE has error free retrieval and fast convergence.

We consider models with neurons as Ising spins. Each neuron exists in two states : $\pm 1$.  A pattern is a configuration  of $N$ binary random variables $\{\xi_1^{\mu},...\xi_N^{\mu}\}$ , where $\mu$ takes $P$ values for $P$ stored patterns. For a configuration $C_N: \{ s_1, s_2, \ldots s_i, \ldots ,s_N\}$, a random variable 
\begin{align}
    m^{\mu} = \frac{1}{N}\sum_{i=1}^{N}\xiu\si
    \label{eq1}
\end{align}
captures the overlap between the $\mu^{th}$ memory and $C_N$. If the configuration $C_N
$ matches with the pattern $\mu$, then $m^{\mu}=1$. We define a $P$ dimensional vector ${\bf m}= \{m^1,m^2.....m^P\}$  to represent the overlap between a configuration and the patterns. 

%The Hamiltonian of the model is built using the Hebb rule \cite{hebb}. 
A general DenseAM \cite{krotov1} is defined by an energy function/Hamiltonian of the form:
\begin{align}
    H(\vv{m}) =-\sum_{\mu=1}^{P} F \left(\sum_{i=1}^N \xi_i^{\mu} s_i \right)   = -N f(\vv{m})
    \label{eq2}
\end{align}
where $F(x)$ is a smooth function. The gradient descent dynamics then works in the direction of lowering the energy and stops when energy can no longer be decreased via single spin flip\cite{krotov2}. Hence $H$ can  be considered the cost function of the memory retrieval which is to be minimized. In this work we develop a formalism using large deviations to get the free energy for any function $f$. 

We consider a set of $P$ randomly chosen quenched patterns represented by the set of $P$ vectors $\{\vv{\xi^{\mu}}\}$, where $\vv{\xi^{\mu}}=\{\xi^{\mu}_1,...,\xi^{\mu}_N\}$. For a random configuration $C_N$ of $N$ neurons, the probability for the configuration overlap to be $\vv{m}$ is proportional to $e^{-\beta H(\vv{m})}$ under  Gibbs  measure, where $\beta$ is the inverse temperature that controls the noise. This probability depends only on the value $\vv{m}$ and is same for all configurations with the same value of $\vv{m}$ for a fixed $\{\vv{\xi^{\mu}}\}$. We hence represent it by $Q_{H,\beta}(\vv{m},\{\vv{\xi^{\mu}}\})$ . This is given by
\begin{align}
Q_{H,\beta}(\vv{m},\{\vv{\xi^{\mu}}\}) = \int_A Q(\vv{m},\{\vv{\xi^{\mu}}\})\exp(N \beta f(\vv{m}))
\label{eq9=10}
\end{align}
where $A$ is the subset of all possible configurations compatible with the value $\vv{m}$ for $P$ patterns. 

The function $Q(\vv{m},\{\vv{\xi^{\mu}}\})$ is the probability distribution of the 
order parameter $(\vv{m})$ for a quenched set of patterns when there is no cost function($H=0$). This is given by 
\begin{align}
     Q(\vv{m},\{\vv{\xi^{\mu}}\})= \sum_{C_N}p(C_N) \prod_{\mu =1}^{P}\delta\left(\sum_{i=1}^{N}\xiu\si= Nm^\mu\right)
    \label{eq3}
\end{align}
here $p(C_N)=p\{s_i\}$ is the probability of occurrence of a configuration $C_N=\{s_i\}$. 
%\textcolor{blue}{For $P=1$ the probability can be evaluated using the double Bernoulli series as shown in SI.}

For a general $P$ we use the integral representation of the delta function by introducing a set of variables $\vv{\lambda} =\{\lambda_1,...\lambda_{\mu}...,\lambda_P\}$. We then get
\begin{align}
Q(\vv{m},\{\vv{\xi^{\mu}}\})&=\int \frac{d^P \vv{\lambda}}{(2\pi)^P} \exp(-N\vv{\mkern1mu\lambda\mkern-1mu}\cdot \vv{m})\,\\\nonumber&\left(\sum_{\{s_i\}}p({\{\si\})} \exp(\sum_\mu \lambda_\mu \sum_i \xiu\si)\right)
\label{eq4}
\end{align}
The sum over spins $s_i$ can be done easily to give 
\begin{align}
&=\int \frac{d^P \vv{\lambda}}{(2\pi)^P} \exp(-N\vv{\mkern1mu\lambda\mkern-1mu}\cdot \vv{m})\prod_{i=1}^{N} \cosh\left(\sum_{\mu}\lambda_\mu \xiu\right)\\   
&= \int \frac{d^P \vv{\lambda}}{(2\pi)^P} \exp\left[-N\left(\vv{\lambda} \cdot \vv{m} - \Psi(\vv{\lambda},\{\vv{\xi^{\mu}}\})\right)\right]
   \label{eq5}
\end{align}
%\begin{align}
 %   G(\vv{\lambda};\vv{\xi}) &= \sum_{{\si}}p({\{\si\})} \exp(\sum_\mu \lambda_\mu \sum_i \xiu\si)\\
%&=\prod_{i=1}^{N} \cosh\left(\sum_{\mu}\lambda_\mu \xiu\right)  \equiv \exp(-N\Psi(\vv{\lambda};\vv{\xi}))
%\end{align}
where $\Psi(\vv{\lambda}, \{\vv{\xi^{\mu}}\}) = \frac1N\sum_{i=1}^{N} \ln(\cosh(\sum_{\mu}\lambda_\mu \xiu)).$ The information about the quenched patterns is contained only in the function $\Psi$. The contours of integration in Eq. \ref{eq5} are understood to be analytically deformed, so that they pass the saddle point. The resultant probability now represented as $Q(\vv{m})$ satisfies large deviation principle with the rate function $I_0(\vv{m})$ , i.e $Q(\vv{m}) \asymp \exp[-NI_0(\vv{m})]$. The saddle point approximation then gives
%In the limit $N \rightarrow \infty$, via the law of large numbers the function is given by
%\begin{align}
%\Psi(\vv{\lambda}) = \Av{\ln\cosh(\sum_{\mu}\lambda_\mu \xi^\mu)},
%\label{eq6}
%\end{align}
%where, $\Av{\cdots}$ is the average over the distribution from which the patterns $\vv{\xi}$ have been drawn. \textcolor{blue}{We hence find that in the thermodynamic limit, $\Psi$ does not depend on the specific patterns but only on the distribution from which the patterns have been drawn. This is true even when $P=1$.} 
%The contours of integration in Eq. \ref{eq5} are understood to be analytically deformed, so that they pass the saddle point. The resultant probability now represented as $Q(\vv{m})$ satisfies large deviation principle with the rate function $I_0(\vv{m})$ , i.e $Q(\vv{m}) \asymp \exp[-NI_0(\vv{m})]$. The saddle point approximation then gives 
\begin{align}    
    I_0(\vv{m}) &= \sup_{\vv{\lambda} \in \mathbb{R}^P} \left[ \vv{\lambda}\cdot \vv{m} - \Psi(\vv{\lambda})\right]
    \label{eq7}
\end{align}
Let $\vv{\lambda}^*$ be the supremum of the function on the right. Then the rate function is $I_0(\vv{m}) = \vv{\lambda}^*\cdot \vv{m} - \Psi(\vv{\lambda}^*)$. 

The $\vv{\lambda^*}$ is a solution of a set of $P$ equations of the kind 
\begin{align}
m^{\mu}_* = \left. \frac{\partial \Psi(\vv{\lambda})}{\partial \lambda_{\mu}} \right\rvert_{\vv{m_*}} 
\label{eq8}
\end{align}
For $P=1$ it is easy to solve the above equation. For $P=1$, $I_0(\vv{m})$ is simply the rate function of a set of $N$ non-interacting Ising spins. In general, in order to find $\vv{\lambda}^*$,  we need to find the solution of the above $P$ dimensional array of non linear equations. This is extremely hard even for $P=2$. For $P>1$ we circumvent the step where the hard inversion need to be performed by developing a procedure similar to the one used by us for random field ferro-magnets \cite{soheli-sumedha, sumedha-barma} which we illustrate below.

%The probability of a configuration $C$ of  neurons under Gibbs measure is proportional to $\exp(-\beta H(\vv{m}))$ with $\beta$ as the inverse temperature that controls the noise. Since $\vv{m}$ is a random variable drawn from a distribution $Q(\vv{m})$ that is defined on $\{-1,1\}^P$,  the probability $Q_{H,\beta}(\vv{m})$,  in the presence of the cost function $f(\vv{m})$ is given by 
%\begin{align}
%Q_{H,\beta}(\vv{m}) = \int_A Q(\vv{m})\exp(N \beta f(\vv{m}))
%\label{eq9=10}
%\end{align}
The rate function $I_{\beta}(\vv{m})$ such that $Q_{H,\beta}(\vv{m} )\asymp \exp(-N I_{\beta}(\vv{m}))$ via the tilted large deviations principle \cite{hollander,sumedha-nabin} is given by
\begin{empheq}{align}
    I_\beta(\vv{m}) &= I_0(\vv{m}) - \beta f(\vv{m}) -\sup_{\vv{x}} \left[ I_0(\vv{x}) - \beta f(\vv{x}) \right]
    \label{eq11}
\end{empheq}
%where $I_0(\vv{m}) = \vv{\lambda}^*\cdot \vv{m} - \Psi(\vv{\lambda}^*)$.
% Assume that, in the thermodynamic limit:
% \begin{align}
%     \lim_{N\to \infty} \frac{\ham_N(\vv{m})}{N} = f(\vv{m})
% \end{align}
The function $I_{\beta}(\vv{m})$ is the free energy landscape for the system whose global minima gives the value of the free energy at a given $\beta$. Let  $\vv{m}_*$ be a fixed point of $I_{\beta}$, then at $\vv{m}_*$, since $\left. \frac{\partial I_{\beta}}{\partial m^{\mu}}\right \rvert_{\vv{m}_{\ast}}=0$, 
\begin{align}
\left.  \frac{\partial I_0}{\partial m^{\mu}}\right\rvert_{\vv{m_{\ast}}}=\beta \left.  \frac{\partial f}{\partial m^{\mu}}\right\rvert_{\vv{m_{\ast}}}
\label{eq12}
\end{align}
%Using $I_0(\vv{m}) = \vv{\lambda^\ast}(\vv{m})\cdot \vv{m} - \Psi(\vv{\lambda^\ast}(\vv{m}))$, this implies,
%\begin{align}
%\pdv*{\left[\vv{\lambda}\cdot \vv{m} - \Psi(\vv{\lambda})\right]}{\lambda_\mu} = 0 
 %   \implies~m^\mu_* =\left  \frac{\partial \Psi(\vv{\lambda})}{\partial \lambda_{\mu}}\right\rvert_{\vv{m_\ast}}
  %  \label{eq12}
%\end{align}
Hence, we have
\begin{align}
\label{eq13}
    \pdv{I_0}{m^\mu} &=  \pdv*{\left[\vv{\lambda^\ast}(\vv{m})\cdot \vv{m} - \Psi(\vv{\lambda^\ast}(\vv{m}))\right]}{m^\mu}\\\nonumber
    &= \lambda^\ast_\mu + \sum_\nu m^\nu  \left. \frac{\partial \lambda_{\nu}}{\partial m^{\mu}} \right\rvert_{\vv{\lambda}^*}  - \sum_\nu \left. \frac{\partial \lambda_{\nu}}{\partial m^{\mu}} \right\rvert_{\vv{\lambda}^*}  \left. \frac{\partial \Psi(\vv{\lambda})}{\partial \lambda^{\nu}} \right\rvert_{\vv{\lambda}^*}
\end{align}

Substituting back in Eq. \ref{eq12}, we get:
\begin{align}
    \lambda^{\ast}_{\mu} = \beta \left. \frac{\partial f}{\partial m^{\mu}}\right \rvert_{\vv{m_{\ast}}}
\label{eq14}
\end{align}
We hence have arrived at the expression of $\vv{\lambda}^*$ without having to invert Eq. \ref{eq8}. This crucial realization allows us to get the expression of the free energy functional for arbitrary associative memory models defined via the function $f$.

We now use the above results for obtaining the free energy functional of the binary neurons. In the limit $N \rightarrow \infty$, via the law of large numbers the function $\Psi$ is given by
\begin{align}
\Psi(\vv{\lambda}) = \Av{\ln\cosh(\sum_{\mu}\lambda_\mu \xi^\mu)},
\label{eq6}
\end{align}
where, $\Av{\cdots}$ is the average over the distribution from which the patterns $\vv{\xi}$ have been drawn. We find that in the thermodynamic limit, $\Psi$ does not depend on the specific patterns but only on the distribution from which the patterns have been drawn.

Substituting $\lambda_{\mu}$ from above into Eq. \ref{eq6} and resulting $\Psi$ into Eq. \ref{eq8} gives the $P$ self-consistency equations for fixed point $\{m^{\mu}_*\}$. They are
\begin{align}
    m_\ast^\mu = \Av{\xi^\mu \tanh(\sum_\nu \beta \left. \frac{\partial f}{\partial m^\nu}\right \rvert_{\vv{m_\ast}}\xi^\nu)}
    \label{eq15}
\end{align}
The rate function, which is the generalized free energy functional of the DenseAMs comes out to be
\begin{align}
\label{eq16}
    I_{H,\beta}(\vv{m}) &=\beta \mathcal{F}(\vv{m})
    =\beta \sum_{\nu}  m^{\nu}_{\ast} \left. \frac{\partial f}{\partial m^{\nu}} \right \rvert_{\vv{m_{\ast}}} \\\nonumber&- \Av{\ln\cosh(\sum_{\nu} \beta \left. \frac{\partial f}{\partial m^{\nu}} \right \rvert_{\vv{m_\ast}}\xi^{\nu})} - \beta f(\vv{m_{\ast}}).
\end{align}
We have defined $\mathcal{F}(\vv{m})= \frac{1}{\beta} I_{H,\beta}$ , as the free energy functional of the system as the free energy is $ \frac{1}{\beta} inf_{\vv{m}} I_{H,\beta}$. The set of $P$ fixed point equations for $\vv{m^*}$ are also given by $\partial \mathcal{F}/\partial m^{\mu}=0$. Hence the resultant $\mathcal{F}$ not just gives the correct position of the fixed points and correct values of the free energy, it also gives the value of all higher derivative of $\mathcal{F}$ at the fixed points. It hence also captures the shape of the landscape, i.e hills and valleys correctly. We use these results to study few models of DenseAMs below.

{\it Dense Associative memory Hopfield Model with polynomial interaction:} The Hamiltonian is 
\begin{align}
    H_N = -\frac{N}{k!}\sum_{\mu = 1}^{P}(m^{\mu})^k
    \label{eq17}
\end{align}
The $f(\vv{m}) = \frac{1}{k!}\sum_{\mu = 1}^{P}(m^\mu)^k$ and $\lambda^\ast_\mu = \frac{\beta}{(k-1)!}(m^\mu)^{k-1}$. Substituting in Eq. \ref{eq15} and \ref{eq16} we get  
\begin{align}
\label{eq18}
\quad m^\mu &= \Av{\xi^\mu \tanh(\sum_\nu \frac{\beta}{(k-1)!}(m^\mu)^{k-1} \xi^\nu)}
\end{align}
\begin{align}
\label{eq19}
\quad \mathcal{F}(\vv{m}) &= \frac{k-1}{k!}\sum_{\mu = 1}^{P}(m^\mu)^k \\\nonumber&- \frac1\beta \Av{\ln\cosh(\sum_\nu \frac{\beta}{(k-1)!}(m^\nu)^{k-1}\xi^\nu)}
\end{align}
Note that while for even $k$, the above equations hold for $-1<m^{\mu}<1$, for odd $k$ they are valid only for positive values $0<m^{\mu}<1$. This is because for odd $k$, negative values of $m^{\mu}$ results in a higher energy than the mirror state with positive $m^{\mu}$. 
%For even $k$ the two states are the two equivalent representations of the same pattern, obtained via flipping each spin. 
The Eq. \ref{eq18} matches  with the stochastic dynamics fixed point equation for DenseAMs obtained recently\cite{rooke}. For $k=2$, the expression of $\mathcal{F}(\vv{m})$ derived above matches with the classic results in \cite{amit1}.

The $\mathcal{F}(\vv{m})$ for $k>2$ however was not known from the earlier studies. We have obtained the exact expression of the free energy functional for any $k$. 
%\sout{The large deviations approach allowed us to handle higher order interactions with ease, which is not possible with the standard Hubbard-Stratonovich transformation.} 
The Eqs. \ref{eq18} and \ref{eq19} can be used to study DenseAM for any $k$ and $P$. 
\begin{figure*}[t]
    \centering
    \includegraphics[width=0.24\textwidth]{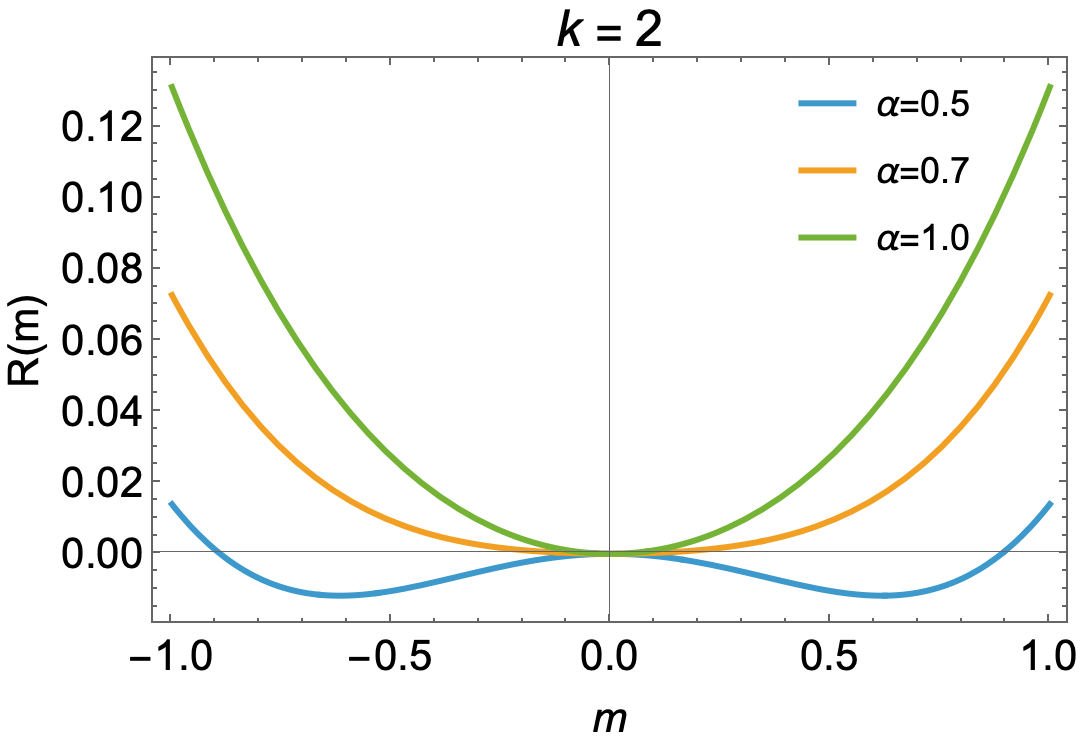}
    \includegraphics[width=0.24\textwidth]{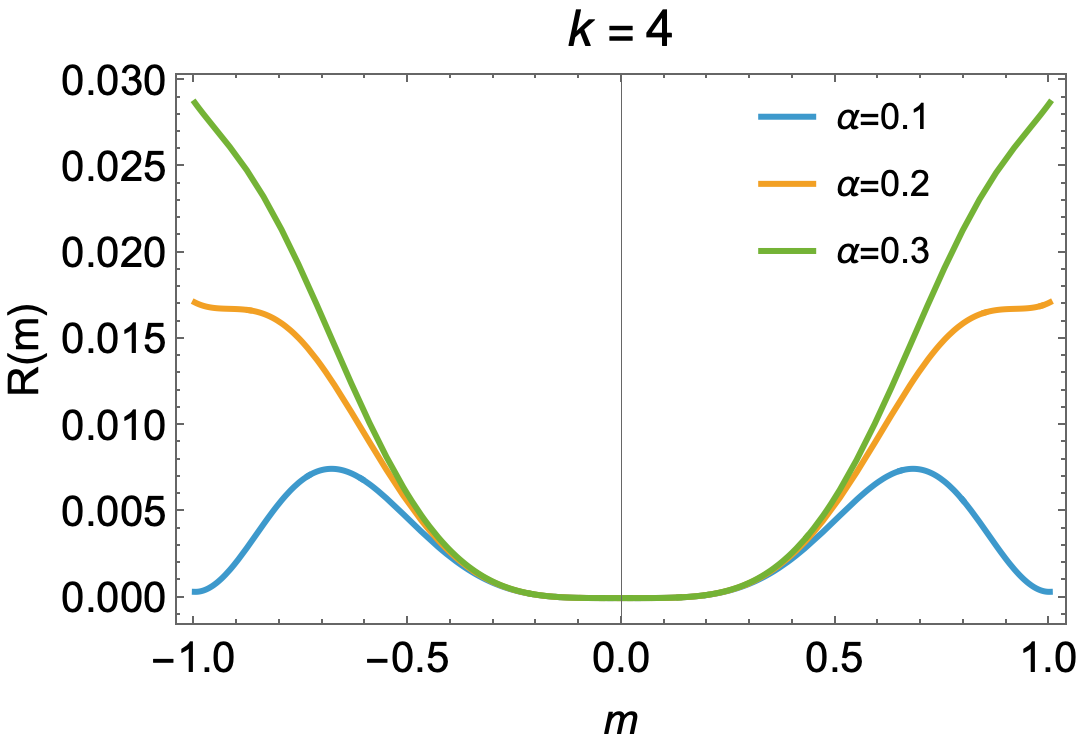}
    \includegraphics[width=0.26\textwidth]{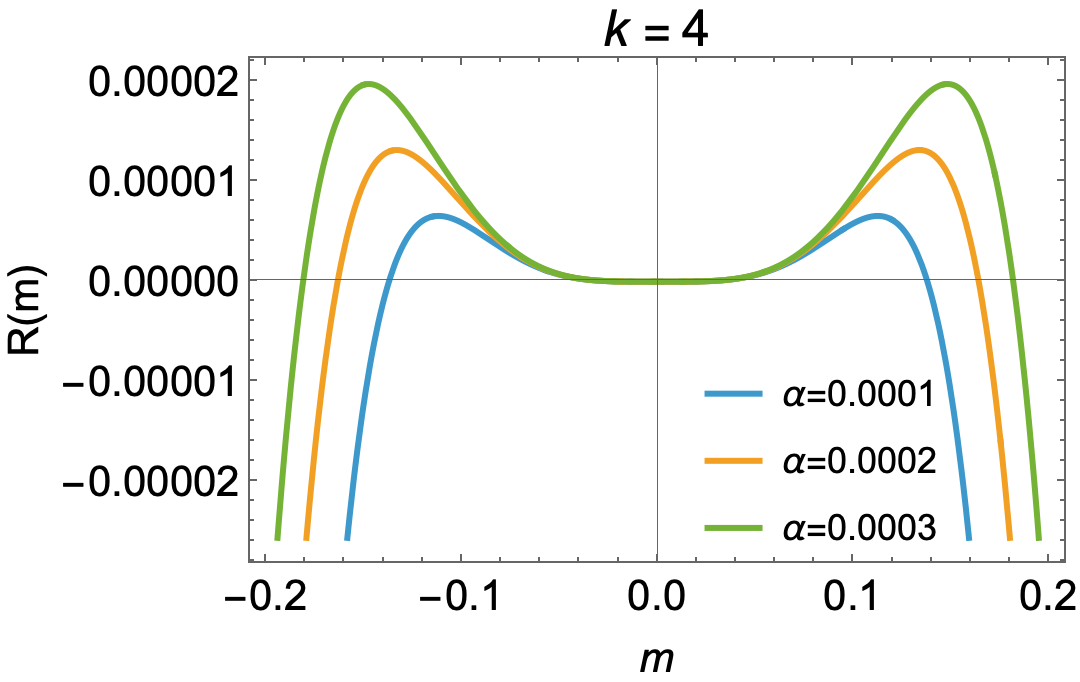}
    \includegraphics[width=0.24\textwidth]{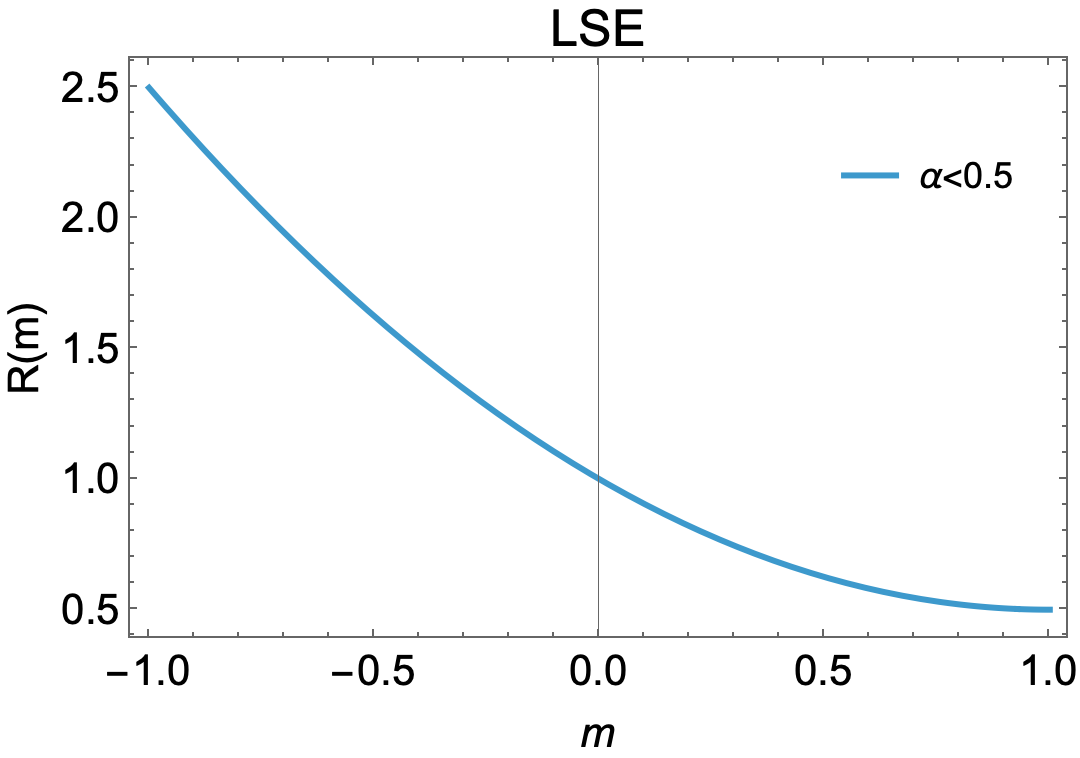}
\caption{Plot of disorder averaged ground state energy of retrieval($R(m)$) of a  random memory when $N,P \rightarrow \infty$. a) for Hopfield model,b) $k=4$ DenseAM. In both cases the function $R(m)$ is plotted near, above and below the transition, In (c) $k=4$ basin of $m=0$ state is shown to illustrate that it is always present. In (d) $R(m)$ for LSE model below and above the transition is shown.}
\label{fig1}
\end{figure*}

For $P=1$, the $\mathcal{F}$ and $m_{\mu}$ equation are similar to that of a $k$-spin Ising model. Hence single pattern DenseAMs has  continuous transition for $k=2$ and first order transition for $k \geq 3$. The transition temperatures for example for $k=2,3, 4$ are $T_c=1/\beta_c=1,0.23,0.04$ respectively. The stability analysis can also be performed straightforwardly by finding the eigenvalues of the Hessian for $P=2$. 
%as shown in the SI. 

The capacity of a network is defined as the maximum number of patterns that can be stored such that a randomly chosen pattern is retrieved with a small error as $N , P \rightarrow  \infty$ at zero temperature. We study the $\mathcal{F}(\vv{m})$ by taking $N \rightarrow \infty$, $P \rightarrow \infty$ along with $\beta \rightarrow \infty$ for the retrieval of a randomly chosen pattern. Each of $(m^{\mu})^k$ corresponding to other patterns is taken as a random variable distributed with mean $0$ and standard deviation $\frac{1}{N^{(k-1)/2}}$ \cite{krotov1}. This is a standard assumption for obtaining the typical behavior as was also used in some recent studies involving modern Hopfield networks \cite{lucibello,rooke}. By central limit theorem, the $z=\sum_{\mu=2}^P (m^{\mu})^{k-1}\xi_{\mu}$  then is a Gaussian random variable with mean $0$ and variance $P/N^{k-1}$. For $P=\alpha N^{k-1}$, $\alpha$ is the variance and $p(z)=\frac{1}{\sqrt{2 \pi \alpha}} e^{-z^2/2 \alpha}$. Taking $m^1=m$ in the thermodynamic limit, we get
\begin{align}
\label{eqxyv}
\quad \mathcal{F}(\vv{m}) &= \frac{k-1}{k!}  m^k \\\nonumber&- \frac1\beta \Avm{\ln\cosh( \frac{\beta}{(k-1)!}(m^{k-1}\xi+z))}
\end{align}
We dropped the term $\sum_{\mu=2}^{P} (m^{\mu})^k$ as its contribution goes to $0$ in the limit $N,P \rightarrow \infty$. The average now is over binary variable $\xi$ and Gaussian random variable $z$. In the limit of $\beta \rightarrow \infty$, this gives the disorder averaged ground state energy of the system as
\begin{align}
\label{eq20}
    R(\vv{m})  &=\frac{k-1}{k!} m^k-\frac{1}{(k-1)!}\sqrt{\frac{2 \alpha}{\pi}} exp\left(\frac{-m^{2 (k-1)}}{2 \alpha}\right)\\\nonumber 
    &-\frac {m^{k-1}}{(k-1)!} \text{erf} \left(\frac{m^{k-1}}{\sqrt{2 \alpha}}\right)
\end{align}
The Eq. \ref{eq20} gives the landscape of the gradient dynamics \cite{sumedha}. The fixed points of Eq. \ref{eq20} are given by
\begin{align}
m = \text{erf}\left(\frac{m^{k-1}}{\sqrt{2 \alpha}}\right)
\label{eq21}
\end{align} 
%For $k=2$ this equation is the \textcolor{blue}{similar} as the fixed point equation for $m$ obtained in \cite{amit2}({\textcolor{red}{add berra ref here}}),  %with $\alpha= r \alpha$, where $r$ is mean square random overlap obtained via replica calculation. 
%In the rest of the paper , we take $e_k=1$ and hence $\alpha=\alpha$.

We next study the consequence of $R(m)$ for DenseAMs in a bit more detail: for $k=2$, the function $R(m)$ has a minima at $m=0$ for high $\alpha$, which continuously changes into a double well as $\alpha$ is lowered as shown in Fig. \ref{fig1}. The system falls into one of the two minima spontaneously with decreasing $\alpha$ at $\alpha_{g}=2/\pi$. The local and global minima are the same and gradient descent always reaches the global minima.

For $k>2$ the system undergoes a first order  transition from $m=0$ to $m \neq 0$ state. The $m = 0$  stops being a global minima at $\alpha_g$ but continues to be a local minima as $\alpha$ is decreased. This can be seen by evaluating the second derivative of $R(m)$($\chi(m)$) at $m=0$. The $\chi(m)$ is given by
\begin{align} 
\chi(m)= 1-(k-1)\sqrt{\frac{2}{\pi \alpha}} m^{k-2} exp \left({\frac{m^{-2 (k-1)}}{ 2 \alpha}}\right)
\label{eqd2}
\end{align}
Equation $\chi(m)=0$ gives the spinodal line and its sign at a fixed point reflects the nature of that fixed point. For $k>2$, $\chi(0)=1$ and hence $m=0$ is a minima for all values of $\alpha$, though it stops being a global minima at a certain $\alpha_g$ (see Fig. \ref{fig1}). As a result, the steady state of gradient descent depends on the initial starting state of the system. If the initial state is in the basin of attraction of $m = 0$, the pattern is not retrieved for any $\alpha$. At a threshold $\alpha_l>\alpha_g$, the $m \neq 0$ minima shows up for the first time. As $k$ increases the basin of attraction of this non zero fixed point shrinks, though the minima gets closer to $m=1$. As a result there is less error in retrieval, provided the  starting state is in the basin of the retrieval state. Since basin reduces, choices of initial state for successful memory retrieval gets more restricted. The Eq. \ref{eq21} gives the location of all the maxima and minima. The size of the basin can hence be estimated by finding the distance to  the closest maxima.

The threshold on $\alpha_k$ for reliable retrieval of memory depends on the percentage of allowed error (given by $(1-m)/2$). For example, for $m=0.99$ we get $\alpha_{th}=0.147$ from Eq. \ref{eq21} that matches well with the threshold derived by Hopfield \cite{hopfield}. As $k$ increases, the error at the transition itself is small and the pattern gets retrieved with higher accuracy. It approaches $1$ as $k \rightarrow \infty$ (see Table \ref{table}). It is hence not surprising that for LSE model with exponential term, the full retrieval is achieved at the transition.
%A bound for $0.5\%$ was obtained in \cite{krotov1} for general $k$. This threshold is different \textcolor{red}{check %how different} than the transition threshold discussed above as the $m$ at the transition though non zero approaches %%%$1$ only in the limit of $k \rightarrow \infty$ (see Table  \ref{table}). We show next for LSE the two thresholds %match completely due to full pattern retrieval at and below the transition.
\begin{table}[]
\begin{tabular}{|c|c|c|c|}
\hline
        $k$ & $\alpha_g$ &  $\alpha_l$ & $m_*$\\\hline
         2 & 0.64 & 0.64 & 0\\\hline
         3 & 0.18 &0.26 & 0.85\\\hline
         4 & 0.1 & 0.2 & 0.92\\\hline
         5 &0.063 & 0.17 & 0.95\\\hline
         10 & 0.015 & 0.13 &  0.98 \\\hline
    \end{tabular}
    \caption{For $\alpha<\alpha_l$ a local minima at $m_*$ in $R(m)$ appears which becomes a global minima at $\alpha_g$. }
    \label{table}
\end{table}
 
{\it Log-Sum-Exponential (LSE) model:} For exponential interaction, the energy function is modeled using a LSE function \cite{ramsauer} as
\begin{align}
   H_N(\vv\eta|\vv{\xi}) = \frac12 ||\vv\eta||^2 - \frac1\lambda \ln(\sum_{\mu =1}^P \exp(\lambda \vv\eta\cdot\vv{\xi^\mu})) 
\label{eq22}
\end{align}
where $\vv{\eta}$ is the state of the neuron which is taken to be continuous and $\lambda$ is the interaction strength. Other exponential cost functions with binary spins using energy dynamics have also been studied\cite{demi,albanese}. We here focus on LSE. 

The second term is the LSE term and is equal to $\frac1\lambda \ln(\sum_{\mu =1}^P \exp(N\lambda m^\mu))$. The LSE term results in an exponential  capacity and fast convergence. The first term is added to ensure a finite norm in the case of continuous states and to bound the energies. Since for binary states the norm is $1$, we add a Hopfield  model like quadratic term to bound the energies. The behavior of the model is still dominated by the LSE function and we are able to reproduce the threshold for retrieval that matches the  threshold derived  in \cite{lucibello}. For binary spins we consider
\begin{align}
   H_N(\vv{m}|\vv{\xi}) = \frac{-N}{2} \sum_{\mu=1}^P  ({m^{\mu}})^2 - \frac1\lambda \ln(\sum_{\mu =1}^P \exp(N\lambda m^{\mu})) 
\label{eq22}
\end{align}
For $P=1$, it is rather trivial and is equal to $H_N(m)= -\frac{N m^2}{2}-m N$. Substituting the $f=-H/N =m+m^2/2$ in Eq. \ref{eq16}, we get  $\beta \mathcal{F}(m)= \frac{\beta m^2}{2}-\ln \cosh [\beta (1+m)]$. This function has a fixed point at non zero $m$ for all values of $\beta$ and as $\beta \rightarrow \infty$, the fixed point is exactly at $m=1$ implying error free retrieval at $T=0$.

For finite $P$ greater than $1$, one can consider either asymmetric Mattis state or symmetric Mattis state \cite{amit1}. For finite $P$, the $f$ and its derivatives are\footnote{Taking $N \rightarrow \infty$ limit for obtaining $f$ and $\partial f/\partial m^\mu$ first and then evaluating $\mathcal{F}$ gives the same expression as when the limit is taken later.}
\begin{align}
f &=  \frac{1}{2} \sum_{\mu=1}^P (m^{\mu})^2+\lim_{N \rightarrow \infty} \frac{1}{N \lambda} \ln \sum_{\mu=1}^{P} \exp(N \lambda m^{\mu})\\
\frac{\partial f}{\partial m^{\mu}} &= m^{\mu} +\lim_{N \rightarrow \infty} \frac{\exp(N \lambda m^{\mu})}{\sum_{\mu=1}^{P} \exp(N \lambda m^{\mu})}   
\end{align}
For asymmetric case, the pattern with the largest overlap dominates and $\frac{\partial f}{\partial m^{1}}=m^1+1$ and $\frac{\partial f}{\partial m^{\mu}}=m^{\mu}~ \forall \mu=2,....,P$. The resultant free energy functional has a single fixed point at $m^1=1$ and $m^{\mu}=0~ \forall \mu=2,..,P$ as $T \rightarrow 0$. Similarly, we can also solve for symmetric case. We get $m^{\mu} =1/P~\forall \mu$ as the fixed point as $T \rightarrow 0$.

To consider the extensive limit in the number of patterns, we define $\phi =\frac{1}{\lambda N} \text{ln}  \sum_{\mu=2}^P \exp(\lambda N m^{\mu})$. Let us first consider only the LSE term. It is now
\begin{align}
f_{LSE} = \frac{1}{N \lambda} \text{ln} \left(e^{\lambda N m^1}+e^{\lambda N \phi}\right) 
\label{eq23}
\end{align}
In the thermodynamic limit $f_{LSE} =  m^1$ if $m^{1} \ge \phi$ and  $f_{LSE}=\phi$ otherwise. In the thermodynamic limit the function $f$ is
\begin{align}
f &= \frac{1}{2} \sum_{\mu=1}^P (m^{\mu})^2 +m^1~if ~ m^1 \ge  \phi \\
f &=  \frac{1}{2} \sum_{\mu=1}^P (m^{\mu})^2 + \phi~if ~ m^1<\phi
\end{align}
When $\phi > m^1$, $\phi$ is also greater than every $m^{\mu}$ and hence $\partial \phi/\partial m^{\mu} =0~\forall \mu$. In this case the model is just the Hopfield model with capacity that grows linearly with $N$.

For $m^1>\phi$ we first evaluate $\phi$ in the limit of $P \rightarrow \infty$. Using the law of large numbers, we can write  
\begin{align}
\phi =\frac{1}{\lambda N} \ln P \langle \text{exp}(\lambda N m^{\mu}) \rangle_{p(m^{\mu})}
\label{eqphi}
\end{align}
where $\langle \rangle _{p(m^{\mu})}$ represents the average over the distribution of $m^{\mu}=m$. 
To determine the maximum capacity for retrieving a randomly chosen pattern (say $\mu=1$) we take the overlap of other patterns to be $O(1/\sqrt{N})$ as was done in \cite{lucibello}. The $m$ here is then a random variable drawn from a Gaussian distribution with mean $0$ and standard deviation $\sigma/\sqrt{N}$. The $p(m) \sim \text{exp}(-N I_0(m))$ with $I_0 = m^2/(2 \sigma^2)$. Taking the saddle point approximation, we get $\langle \text{exp}(\lambda N m )\rangle _{p(m)} \sim \text{exp}(N \lambda^2 \sigma^2/2)$(at the saddle point $m=\lambda \sigma^2$).
%\sout{Using the tilted LDP, we then get, $\langle \text{exp}(\lambda N m^{\mu} )\rangle _{p(m^{\mu})} \sim \text{exp}(-N \lambda^2 \sigma^2/2)$}. 

For $P$ slower than exponential, Eq. \ref{eqphi}  gives $\phi= \lambda \sigma^2/2$ to leading order in $N$. Thus $\phi$  becomes independent of the number of patterns in the thermodynamic limit and hence $m^1=1$ is always possible by adjusting $\lambda$. Hence $P$ has to grow exponentially for the possibility of retrieval failure. 

For $P \sim \text{exp}(\alpha N)$, we get
\begin{align}
    \phi =\frac{1}{\lambda} \left(\alpha+\frac{\lambda^2 \sigma^2}{2}\right)
    \label{eq24}
\end{align}
The extrema  of $\phi$ occurs at  $\lambda^{*} =\sqrt{2 \alpha}/\sigma$.  For $\lambda <\lambda^*$, $\phi$ increases rapidly with increasing $\lambda$ and is a slow increasing function for $\lambda > \lambda^*$. The $\lambda^{*}$ hence separates the annealed and frozen branch of the model with $m = \lambda^* \sigma^2=  \sigma \sqrt{2 \alpha}$ and $\phi=\sigma \sqrt{2 \sigma}$ in the frozen branch.

For $m^{1} \ge \phi$  the  model behaves similar to  asymmetric case we discussed earlier with
%\begin{align}
%    f_{\phi<m^1} = \frac{1}{2} \sum_{\mu=1}^P {m^1}^2 + m^1
%\label{eq25}
%\end{align}
\begin{align}
\mathcal{F}_{\phi<m^1}(\vv{m}) &= \frac{1}{2} (m^{1})^2 - \frac1\beta\mathbb{E}_\xi\left[\ln\cosh(\beta (m^{1}+1) \xi^{1})\right]
\label{eq28}
\end{align}
%\sout{The $f_{LSE}=-m^1 $ for $\phi < m^1$ and $f_{LSE}=-\phi$  when  $\phi >m^1$. For $\phi>m^{1}$ there is no retrieval. So let us consider the $\phi<m^{1}$ scenario. For binary variables the norm is $1$. But to consider the more general case, we take $||\vv{\sigma}||^2 = N (m^1)^2$ (assuming that $m^{1}$ depends on $\vv{\sigma}$ through its rescaled norm), the $f$ is }
The $\sum_{\mu=2}^P {(m^{\mu})^2}$ does not appear in the above equation as it does not contribute in the thermodynamic limit.
%via the law of large numbers it is proportional of $\langle m^2 \rangle_{p(m)} =0$.
%For $\phi < m^1$ the model behaves just like the standard Hopfield model. Since $P \propto exp(\alpha N)$ and the capacity of Hopfield model is $O(N)$, there is no retrieval in this case. The memory retrieval is possible only for $\phi<m^1$. 

For $\beta \rightarrow \infty$, we get
%\begin{align}
%\label{eq29}
 %  R_{\phi<m^1}(m)&= \frac{1}{2} {(m^{1})}^2-m^1-\sqrt{\frac{2 s }{\pi}}%~\text{exp}\left(-\frac{(m^1-1)^2}{2 s}\right)\\\nonumber&-(m^1-1) \text{erf}%%\left(\frac{m^1-1}{\sqrt{2 s}}\right)
%\end{align}
%with
%\begin{align}
 %   m^1 =1+\text{erf}\left(\frac{m^1-1}{\sqrt{2 s}}\right)
  %  \label{eq30}
%\end{align}
%where $s \propto \sqrt{P}$. Taking the $s  \rightarrow \infty$ limit for %$P\propto exp(\alpha N)$, we get
\begin{align}
   R_{\phi<m^1}(m^1)= \frac{1}{2} (m^{1})^2 -(m^1+1)
   \label{eq31}
\end{align}
which has a fixed point at $m^1_*=1$. This is the free energy landscape of the retrieval phase of LSE. As shown in Fig. \ref{fig1}, the landscape  has tilted towards the $m=1$ state and  the retrieval of the memory is error free. Equating $\phi=1$ then gives us the threshold $\alpha_c(\lambda)$ below which the pattern is completely retrieved with no error. 
We fix $\sigma=1$. Then $\alpha_c(\lambda)\leq 1/2$ for real $\lambda$. Since $\lambda=1$ at $\alpha_c=1/2$. For $\lambda \leq \lambda^*=1$, we get   
\begin{align}
\alpha_c = \lambda \left(1-\frac{\lambda}{2}\right)
\label{eq32}
\end{align}

We have hence reproduced the threshold for retrieval derived recently using the random energy model\cite{lucibello}. It is interesting that threshold is the same in the case of binary and continuous variables. This is because the sum of exponential terms in LSE dominates the behaviour of the model. This threshold separates the annealed branch from the frozen branch. The frozen branch occurs for $\lambda > \lambda^*=1$ with $\alpha$ plateauing at $\alpha_c=1/2$. However to get the estimate of basins of attraction, we need to consider continuous variables as it is trivial with binary variables.

{\it Discussion:} 
%\textcolor{blue}{Our method is general method which cannot evaluate the details like the value of the overlap as determined by replica method for Hopfield and k-spin interaction model. The strength off our treatment though is its generality and eas with which it can produce the shape of the landscape and the threshold, besides being exact for finite P}.
Our method can evaluate any cost function defined by Eq. \ref{eq2}. Although currently demonstrated using binary variables, the framework generalizes effortlessly to other discrete or continuous variables. We validated our approach by analyzing two prominent AM models. We showed that the disorder-averaged ground state energy  functional $R(m)$, is useful for decoding gradient dynamics. 

In the extensive pattern limit the usual method is the replica trick to evaluate the quenched average. It is especially useful to evaluate the random pattern overlaps\cite{amit2,gardner,albanese1}.  We do not attempt that here. Instead we provide general scheme for analyzing the typical behavior. Interestingly, we find that the fixed point rapidly shifts towards $m=1$  for higher $k$ DenseAMs and LSE model and our calculation gives a good estimate of the critical $\alpha$ for retrieval.

For DenseAMs, we showed a strong initial-state dependence during memory retrieval, as the basin at $(m=0)$ persists for all $\alpha$. This suggests that  a mechanism like a stochastic noise \cite{rooke}  can be useful for retrieval when the initial condition flows to $m=0$ fixed point.

%On the otherhand the formalism presented here using large deviations theory although does not characterize the glassy regime oro calculates the random pattern overlaps,  it provides a characterization of the retrieval basin geometry that gives insights into the dynamical accessibility of the target memory state. For example, we show that the LSE cost function essentially changes the nature of landscape compared to polynomial DenseAMs by tilting the landscape. As a result, LSE has error free retrieval and fast convergence.}
%\textcolor{blue}{The variance of the random overlaps, can be calculated more precisely using replica method as is done in \cite{amit2,gardner,albanese1}. We do not attempt that here. Instead we provide general scheme for analyzing the typical behavior. Interestingly, we find that the fixed point rapidly shifts towards $m=1$  for higher $k$ DenseAMs and LSE model and our calculation gives a good estimate of the critical $\alpha$ for retrieval.} 
%For DenseAMs, we showed a strong initial-state dependence during memory retrieval, as the basin at $(m=0)$ persists for all $\alpha$. This suggests that  a mechanism like a stochastic noise \cite{rooke}  can be useful for retrieval when the initial condition flows to $m=0$ fixed point.

%In the limit of $P \rightarrow \infty$, replica calculations have been performed. We focus on $T=0$ behaviour when one is near a particular memory. This particular scenario can be studied from Eq. and Eq, . This is useful....

%Finally, for LSE, our results align with the findings by Lucibello et al. \cite{lucibello}.
{\it{Acknowledgement:}} We thank the anonymous referee for helpful comments.

\end{document}